# Polarization-Sensitive Au-TiO$_2$ Nanopillars for Tailored Photocatalytic Activity

*Ning Lyu [a b]\*, Anjalie Edirisooriya [b], Zelio Fusco [b], Dawei Liu [c d], Lan Fu [c], Fiona J. Beck [b]\*, Christin David [a e]\**

a. Institute of Solid State Theory and Optics, Friedrich-Schiller-Universität Jena, 07743 Jena, Germany

b. School of Engineering, Australian National University, ACT 2601, Australia

c. Australian Research Council Centre of Excellence for Transformative Meta-Optical Systems, Department of Electronic Materials Engineering, Research School of Physics, The Australian National University, Canberra ACT 2601, Australia

d. Institute of Applied Physics, Abbe Centre of Photonics, Friedrich Schiller University Jena, Albert-Einstein-Str. 15, Jena 07745, Germany

e. University of Applied Sciences Landshut, Am Lurzenhof 1, 84036 Landshut, Germany

\* Corresponding authors: christin.david@uni-jena.de (C. David); fiona.beck@anu.edu.au (F.J. Beck); ning.lyu@uni-jena.de (N. Lyu)

**ABSTRACT**

Plasmonic metasurfaces play a crucial role in resonance-driven photocatalytic reactions by effectively enhancing reactivity *via* localized surface plasmon resonances. Catalytic activity can be selectively modulated by tuning the strength of plasmonic resonances through two primary non-thermal mechanisms: near-field enhancement and hot carrier injection, which govern the population of energetic carrier excited or injected into unoccupied molecular orbitals. We developed a set of polarization-sensitive metasurfaces consisting of elliptical Au-$TiO_2$ nanopillars, specifically designed to plasmonically modulate the reactivity of a model reaction: the photocatalytic degradation of methylene blue. Surface-enhanced Raman spectroscopy reveals a polarization-dependent reaction yield in real-time, modulating from 4.7 (transverse electric polarization) to 9.98 (transverse magnetic polarization) in 10 s period, as quantified by the integrated area of the 480 $cm^{-1}$ Raman peak and correlated with enhanced absorption at 633 nm. The single metasurface configuration enables continuous tuning of photocatalytic reactivity *via* active control of plasmonic resonance strength, as evidenced by the positive correlation between measured absorption and product yield. This dynamic approach provides a route to selectively enhance or suppress resonance-driven reactions, which can be further leveraged to achieve selectivity in multibranch reactions, guiding product yields toward desired outcomes.

Optical metasurfaces have garnered significant attention as a promising platform for artificial photosynthesis, particularly when integrated with active plasmonic and hybrid materials.[1–5] These sub-wavelength structures support localized optical resonances that concentrate light at their surfaces, effectively generating energetic carriers or photons to drive photocatalytic reactions.[6–9] Their optical properties are strongly dependent on their geometric configurations, making them highly tunable—an advantageous feature for controlling photocatalytic reactivity[10]. Morphological engineering of metasurfaces is, thus, a critical strategy for tailoring their resonant behavior, enabling precise control over resonance wavelength and intensity.[11,12]

In the context of artificial photosynthesis, it offer the means to address global decarbonization goals and facilitate the synthesis of renewable fuels,[13,14] especially in hard-to-decarbonize sectors, such as shipping and mining industry. However, the synthesis involves complex, multi-step reaction pathways with numerous intermediates, such as the process *via* converting $CO_2$ into energy-dense hydrocarbons[14,15] or biofuel synthesis of advanced alcohols and esters.[16,17] Recent studies have demonstrated the potential of metasurfaces to modulate these reaction pathways,[18] thereby enhancing catalytic activity[19] and improving product selectivity, an area of growing importance in sustainable energy conversion and carbon mitigation research.[20,21] Especially, as active elements, plasmonic material has been demonstrated to effectively accelerate photochemical reaction and promise of achieving reaction selectivity.[13] With precise morphological design, the optical properties of metasurfaces can be finely tuned to favor specific pathway reactions while suppressing undesirable byproducts, thereby enabling high selectivity in complex, multibranch photocatalytic systems.[22–24]

Plasmonic hybrid metasurfaces, catalyze resonance-driven reactions through localized surface plasmon resonances (LSPR), which enable their catalytic activity.[25] In this process, incident electromagnetic (EM) waves at specific wavelengths excite collective oscillations of conduction electrons within the metallic nanostructures, generating regions of intense electromagnetic field enhancement when resonant, commonly referred to as *"hot spots"* at the metal surface. During the subsequent decay processes, two primary non-thermal charge-transfer mechanisms can occur which can initiate or activate photochemical reactions.26 First, photon absorption in the near-field region can promote electrons from the highest occupied molecular orbital (HOMO) to the lowest unoccupied molecular orbital (LUMO) within the adsorbed molecules, thus facilitating chemical transformations (Figure 1b). Second, hot electrons generated from LSPR decay can be injected into molecular orbitals either directly with hybridized surface states (as illustrated by the dotted line in Figure 1c) or indirectly with the energetic carrier diffusion (solid line in Figure 1c), potentially activating and enhancing the reaction.27,28 If these energetic carriers are not transferred to the reactant molecules, they undergo relaxation *via* electron-electron scattering, leading to localized heating of the nanostructure. This photothermal effect can also contribute to enhancing reaction rates, particularly in thermally assisted catalytic processes.29 The two primary non-thermal mechanisms of plasmonic photocatalysis, near-field enhancement and hot carrier transfer, enable the tailorable activation of specific chemical bonds, facilitating the catalytic reaction process.30

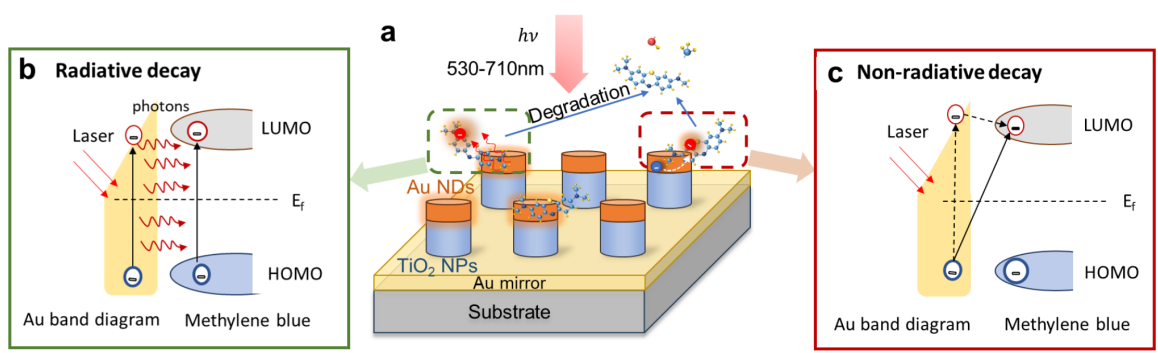

**Figure** 1: Schematic of an Au-TiO$_2$ elliptical nanopillar array (a) for methylene blue (MB) degradation reactions, illustrating two non-thermal LSPR effects: Radiative decay (b) and non-radiative decay (c) with direct carrier injection.

Various metasurface designs have been applied in light-driven catalytic reactions, with morphological engineering playing a key role in enhancing their performance. For example, Dutta *et al.* designed a sandwiched gold-hematite-gold nanodisk structure that extends the effective wavelength range for photoelectron generation beyond the intrinsic bandgap of hematite, facilitating the plasmon-induced generation of hot electrons at near-infrared ranges.31 Morphology tailoring of metasurfaces also shows significant promise for enhancing specific photocatalytic reactions by supporting resonant modes that match the spectral characteristics of target processes. Hu *et al.* developed a $TiO_2$-based metasurface relying on optical bound states in the continuum (BIC), demonstrating broad spectral tunability. By adjusting the scaling factor, tilting angle, and material composition of the nanostructure, they significantly boosted photocatalytic activity for $Ag^+$ reduction.22

Furthermore, metasurface designs with high tunability can enable active control over reaction rates. For instance, Yuan *et al.* employed $Si_3N_4$ nanocubes coated with Ni nanoparticles that support polarization-dependent quasi-BIC modes, modulating the reactivity of the $H_2$ dissociation reaction.5 In our group, we systematically investigated the control of reactivity in a model photocatalytic reaction using a series of Au nanoparticles coupled with $TiO_2$ nanocavities. By tuning the plasmonic resonance strength through modulation of the Fabry-Pérot resonance peak across different samples, we were able to control the resulting product yield.32 However, the variations in samples affect the plasmonic modulation of light-driven reactions, due to the various Au absorption proportion and fabrication defects. To improve flexibility and potential in precision chemistry, it is desirable to achieve yield manipulation within a single, compact, and integrated device. Metasurfaces are employed to selectively drive resonance-enhanced reactions through precisely designed configurational features. Thereby, in this work, we focus on polarization-dependent metasurfaces which offer an effective strategy for continuous controlling optical responses and enabling tailorable photocatalysis within a single metasurface design.

We experimentally investigate the optical tunability of elliptical Au-TiO$_2$ nanopillars (Au-TiO$_2$ NPs) under various light polarizations, focusing on demonstrating their application in controlling resonance-driven catalytic reactions using the N-demethylation of methylene blue (MB) as a model reaction. The asymmetry in the nanopillars gives rise to polarization-dependent plasmonic resonances, which are tuned to the effective wavelength for this model reaction. The catalytic performance is attributed to the two aforementioned non-thermal effects of LSPR: near-field enhancement and hot carrier transfer. By modulating the polarization of the incident light, we demonstrate active control of the catalytic activity through polarization-induced tunability of the plasmonic resonance in a single metasurface design based on elliptical Au nanodisks on top of TiO$_2$ nanopillars.

RESULTS AND DISCUSSION

**Metasurface Design**

We designed an array of Au-TiO$_2$ nanopillars, as illustrated in Figure 1a, where each TiO$_2$ nanopillar is capped with a gold nanodisk (Au ND). Both the nanopillars and nanodisks feature elliptical geometries with independently defined major and minor axes ($R_x$ and $R_y$), enabling the tuning of the resonance wavelengths under two orthogonal polarizations of incident light. This polarization-dependent optical behavior is central to achieving tunability and enabling shifts in the product yield of resonance-driven reactions. Proper tuning of the resonant behavior can be leveraged in multibranch photocatalytic reactions, to achieve enhancement of desired reaction pathways while simultaneously suppressing undesired byproduct formation.

We target the metasurface resonance to overlap with the model reaction: the N-demethylation of methylene blue (MB) (see Figure S1). The photocatalytic performance of light driven reactions is strongly wavelength-dependent, governed by the HOMO-LUMO energy gap of the reactant molecules. For MB, the active range lies between about 530 - 710 nm.[25]

When the plasmonic resonance of the metasurface aligns with the molecular electronic transitions, photons and hot electrons are efficiently injected into the LUMO of MB, significantly enhancing the reaction rate. In contrast, when the resonance peak is detuned from the HOMO-LUMO gap, the localized electric field strength is diminished at the effective wavelength. This misalignment reduces the effectiveness of both near-field enhancement and hot carrier transfer, achieving suppression of the catalytic activity.

The elliptical Au-$TiO_2$ NPs consisted of $TiO_2$ NPs of 100 nm height, each capped with a 15 nm thick Au ND, positioned on the gold mirror substrate. The optical properties of the elliptical Au-$TiO_2$ NP metasurface were simulated using *COMSOL Multiphysics* under two perpendicular polarization states: transverse magnetic (TM) and transverse electric (TE). To systematically investigate the effect of structural asymmetry, we varied the nanopillar radius along the y-direction ($R_y$) from 30 nm to 70 nm while keeping the x-direction radius ($R_x$) fixed at 35 nm.

The plasmonic resonance, primarily determined by the Au ND radius, exhibits polarization-dependent variations under switching incident light conditions,[12] as shown in Figure 2a and Figure S2 in the supporting information (SI). The peak absorption wavelength redshifts from 620 nm to 950 nm when illuminated by linear polarized light aligned with the $R_y$ direction (TE mode). In contrast, under TM polarization, the resonance peak remains relatively stable at 630 - 650 nm. Considering the available laser wavelength for the characterization, we target 633 nm for the morphological design. In the design with $R_y$ = 60 nm, the absorption at the reaction target wavelength increased from 27.9% to 99.9% switching from TM to TE excitation (see Figure 2a). This polarization-dependent absorption behavior can be leveraged for tunable metasurface designs.

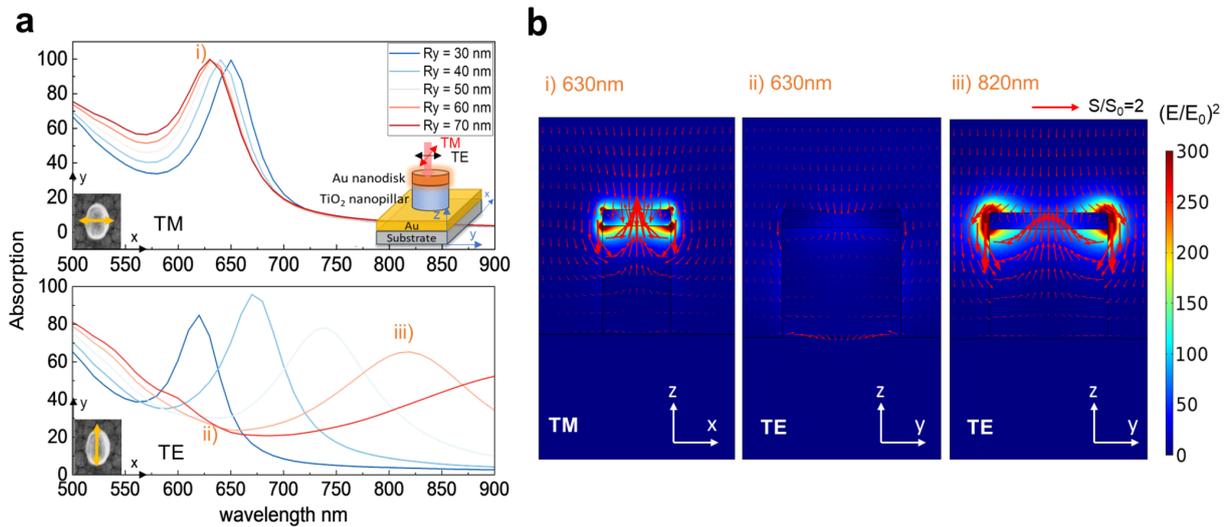

**Figure** 2: Simulation of Au-TiO$_2$ NP arrays with an increasing radius in the y-direction: a. Simulated total absorption spectra for varying R$_y$ from 30 to 70 nm, and R$_x$ kept at 30 nm. The incident light is linearly polarized in TM (top) and TE (bottom). The insets (bottom left) illustrate the direction of polarization. The inset at bottom right shows the 3D model of a single Au-TiO$_2$ NP unit cell and the definitions of axes. b. Electric field enhancement (color scale) and normalized Poynting vectors (red arrows) for a unit cell of the nanopillar structure at i) R$_y$ = 60 nm under TM polarization at 630 nm (xz plane) and TE polarized light at ii) 630 nm and iii) 850 nm (yz plane). Electric field enhancement is calculated as the square of the electric field strength E relative to the incident field E$_0$. The normalized Poynting vectors **S** represent the time-averaged power flow, normalized by the incident power **S$_0$**.

To investigate the near-field distribution, we present a cross-sectional map of the electric field enhancement $E^2/E_0^2$ of the Au-TiO$_2$ NP unit cell in Figure 2b. Arrows in the figure indicate the time-averaged Poynting vector ***S***, normalized to the incident Poynting vector ***S$_0$***. For the metasurface with R$_y$ = 60 nm, Figure 2b(i) shows the electric field enhancement at the peak wavelength of 630 nm under TM polarization. Figure 2b(ii) and 2b(iii) display the field enhancement at 630 nm (absorption dip) and 820 nm (absorption peak), respectively, under TE polarization.

Significantly enhanced localized electric fields are observed at the absorption peak wavelengths for both TE and TM modes, concentrated near the Au nanodisks due to strong plasmonic effects. Under TM polarization at 630 nm, the maximum electric field enhancement reaches $E_{TM\ 630nm}^z / E_0^z = 1517$ at the surface of the Au nanodisk. In TE mode, the peak shifts to 820 nm due to the larger disk radius along the y-direction, reaching an enhancement of $E_{TE\ 820nm}^z / E_0^z = 1768$. Correspondingly, the Poynting vector plots reveal concentrated energy flow near the Au nanodisks, indicating a strong plasmonic resonance.12 In contrast, at 630 nm under TE polarization, corresponding to a spectral absorption dip, the electric field enhancement is significantly reduced to $E_{TE\ 630nm}^z / E_0^z = 33$, nearly 1/50th of the value under TM polarization at the same wavelength. The corresponding Poynting vector distribution shows much weaker energy flow, with partial propagation into the mirror layer and substrate, indicating diminished resonance within the nanodisk. These results demonstrate that the metasurface is designed to tune the peak plasmonic resonance by varying the polarization state, enabling wavelength-dependent field enhancement. This tunability is key for applications such as photocatalytic reactions, where efficient light-matter interaction at target wavelengths is crucial.

To effectively manipulate reactions by tuning the resonance strength, we optimized the design of Au-TiO$_2$ NPs to maximize the absorption difference between the two perpendicular polarization states at the target wavelength of 633 nm, which lies within the HOMO-LUMO range of the MB reactant molecule, and is compatible with an available laser for characterization. The absorption difference A$_{diff}$ was calculated using the absolute values for both TM and TE modes. Figure 3a shows that the maximum absorption difference occurs when $R_x$ = 35 nm and $R_y$ = 60 nm reaching over 60%, providing the largest enhancement in polarization-sensitivity.

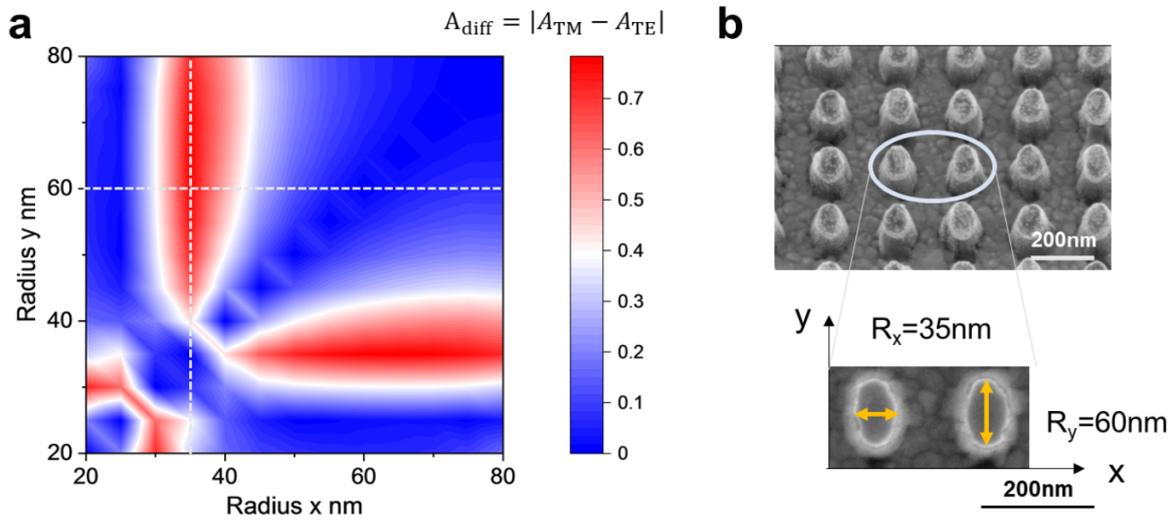

**Figure 3**: a. Absorption difference $A_{diff}$ as a function of nanopillar radius in the x- and y-directions, representing the difference in absorption when illuminating with TM and TE light. The maximum value is observed $R_x$=35 nm and $R_y$=60 nm. These radii are used as the design parameters for fabricated samples. b. SEM images of fabricated Au-TiO$_2$ NPs with the optimized design parameters tilted by 30° and zooming in on two NPs in a top-view image.

**Characterization of Metasurfaces**

A scanning electron microscope (SEM) micrograph image of a representative Au-TiO$_2$ NP metasurface is shown in Figure 3b, fabricated using a top-down lithographic approach, outlined in Figure S3 and Method section. The asymmetric elliptical shape of the nanopillars is evident in the top-view image, with measured radius of $R_x = 35 \pm 3$ nm and $R_y = 60 \pm 4$ nm (determined with 20 measurements). The radius in the x-direction corresponds to the resonance mode when excited by TM polarized light, while the radius in the y-direction supports the resonance for the TE mode.

This fabrication process results in TiO$_2$ NPs exhibiting noticeable tapering in both the x- and y-directions (Figure 3b). The $R_x$ decreases from 50 nm to 35 nm, going from NP's base to top, and similarly, $R_y$ decreases from 74 nm to 60 nm. To evaluate the effect of this geometric deviation, we performed simulations of the total absorption spectrum, as shown in Figure S4. The tapering defect was modelled symmetrically in both directions while preserving the same expand magnitude, such that $R_{xtop} - R_{xbottom} = R_{ytop} - R_{ybottom}$. The results indicate that the total absorption spectrum undergoes a slight redshift (< 5 nm) as $R_{xbottom}$ increases to 57 nm. As discussed in the simulated design results, the total absorption of the metasurface is primarily governed by the plasmonic resonance of the Au NDs. Consequently, the tapering of the $TiO_2$ NPs has a negligible impact on the resonance peak wavelength. However, we observed an increase of the full width at half maximum (FWHM) of the plasmonic resonance peak with the bottom radius, indicating a taper-induced broadening effect on the resonance.

Polarization-dependent absorption is governed by the elliptical dimensions of the Au-$TiO_2$ NPs, specifically the radius in the x- and y-directions.[12] Figure S6 shows the total absorption for samples fabricated with varying electron-beam lithography (EBL) doses (SEM image shown in Figure S5), which incrementally increase the nanopillar dimensions. In both TE and TM modes, increasing the nanopillar size results in a redshift of the absorption peak. For the metasurface design with $R_x$ = 35 nm and $R_y$ = 60 nm (in Figure 4a), the absorption spectrum under TM polarized illumination (dark red curve) exhibits a peak near 650 nm with an absorption of up to 97%. In contrast, under TE polarized light, the peak redshifts to approximately 850 nm due to the larger radius in the y-direction. These results confirm that the Au-$TiO_2$ NP arrays support two distinct resonances corresponding to the x- and y-components of the incident electric field. As the polarization angle is varied, the resonance at 650 nm diminishes while a second resonance at a longer wavelength becomes more prominent under TE polarization. At the target reaction wavelength of 633 nm, the absorption is tunable from 89% to 58% depending on the polarization angle. Consistent with Malus' law, the rate of change in absorption is most significant near a polarization angle of 45° and decreases as the angle approaches 0° or 90°.

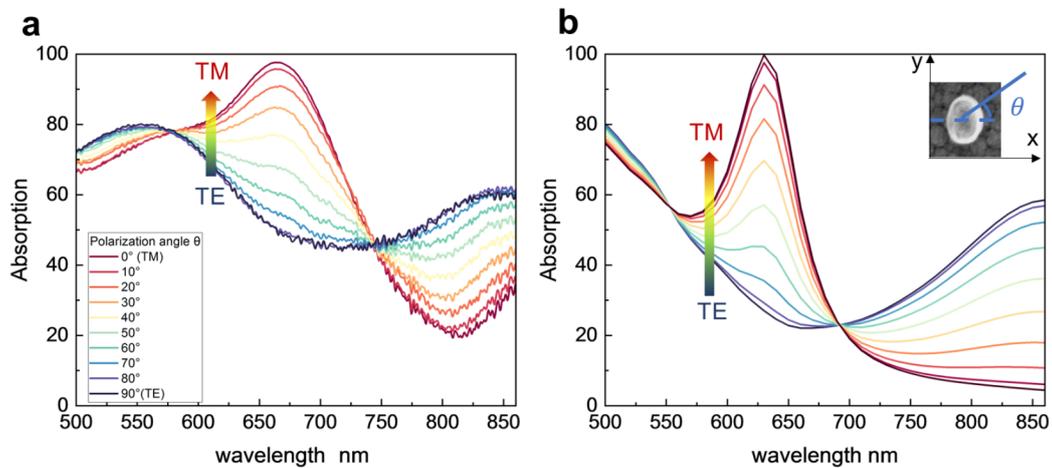

**Figure** 4: Measured (a) and simulated (b) absorption spectrum for Au-TiO$_2$ NP arrays with R$_x$ = 35 nm and R$_y$ = 60 nm varying the polarization direction. The angle of polarization θ is defined as the angle with the x-axis, which shifts from 0° to 90°, as shown in the inset.

A similar trend is observed in the simulated results for the Au-TiO$_2$ NP array model with dimensions R$_x$ = 35 nm and R$_y$ = 60 nm. The absorption at 633 nm exhibits a strong dependence on the polarization state of the incident light. This behavior is primarily attributed to plasmonic resonances in the Au NDs, as illustrated in Figure S2 in the SI. Under TM polarization, the absorption in the Au NDs reaches 96%, accounting for approximately 99% of the total absorption. A slight spectral shift of the absorption peak, from 630 nm in simulation to 650 nm in the measured results, is attributed to fabrication-induced dimensional uncertainties. Furthermore, the FWHM is broader in the experimental data compared to simulation results, which can be explained by the tapering of TiO$_2$ NPs and the finite array size in the fabricated samples, as opposed to the infinite periodicity assumed in the simulations. Nevertheless, at the target wavelength, the polarization-dependent absorption demonstrates a tunability exceeding 30%, predominantly enabled by the plasmonic contribution of the Au NDs.

Furthermore, at the TM peak wavelength of 650 nm for Au-TiO$_2$ NP arrays, the tunability of absorption increases significantly from 52% (TE) to 97% (TM) offering greater flexibility in tailoring optical properties. Similarly, at the peak wavelength under the TE mode, the absorption varies from 61% (TE) to 27% (TM). The maximum absorption occurs under polarization perpendicular to that of the maximum at 650 nm, and *vice versa*. These characteristics demonstrate strong potential for further applications in tuning catalytic reactivity.

**Polarization-dependent photocatalysis**

In the photocatalysis of the model reaction, MB molecules were adsorbed onto the metasurface, where light-induced cleavage of the C-N bond leads to the formation of the reaction products, N-demethylated derivatives.[25] The two primary non-thermal mechanisms of plasmonic photo-catalysis, near-field enhancement and hot carrier transfer, enable the tunable activation of specific chemical bonds, facilitating the catalytic reaction process.[30] The reaction progress can be monitored in real time using surface-enhanced Raman spectroscopy (SERS), based on the specific peaks of the reaction product.

The reaction product was identified and quantified by tracking its characteristic Raman peak[33] at approximately 480 cm$^{-1}$. Detailed experimental procedures are provided in our previous publications.[25,32] The normalized Raman spectra of the MB photocatalytic reaction illuminated under the TM mode are presented in the inset of Figure 5b. The spectra under other polarizations are shown in Figure S8. The increasing intensity of the product peak over time reflects the generation rate of N-demethylated derivatives, which was further analyzed through deconvolution fitting as shown in Figure S9. The integrated area under the fitted curve was used as a quantitative measure of the MB photodegradation yield.

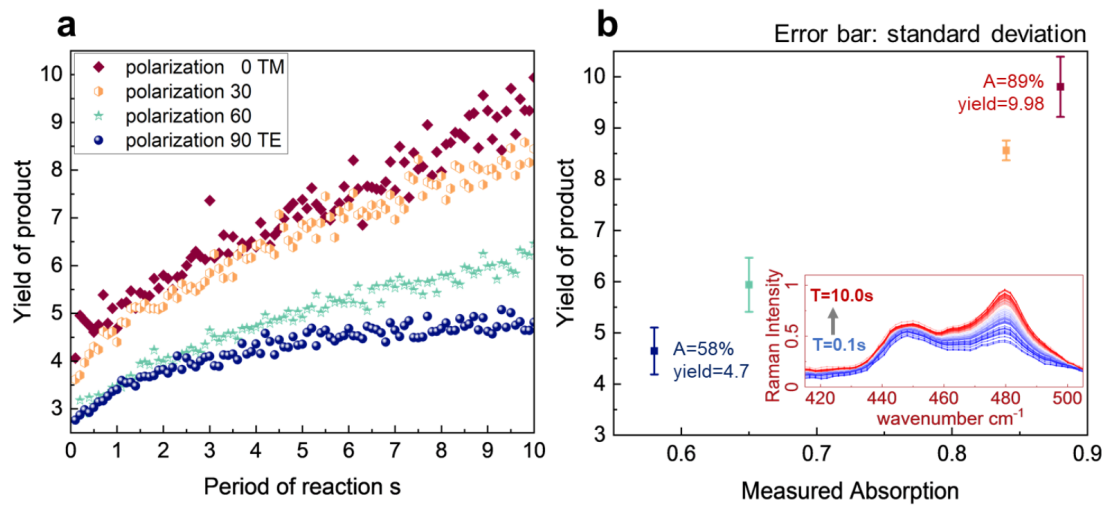

**Figure** 5: a. Yield of the Au-TiO$_2$ NP arrays against the reaction processing time for four different polarization states. b. Averaged yield at the final timestep (t = 10 s) put in relation with the measured absorption. The error bar indicates the standard deviation for five measurements. The inset shows the normalized Raman spectrum under TM mode from 0.1s (blue curve) to 10s (red curve) reaction period.

Figure 5a shows the time evolution of product yield calculated at different reaction time, revealing the photocatalytic dynamics under four distinct polarization states. Under TE polarization (represented by the dark blue dotted curve), the yield exhibits a rapid increase during the initial 3 seconds, rising from 2.76 (at 0.1 s) and stabilizing at approximately 4.7 by 10 seconds. Upon switching the polarization to the perpendicular TM mode, the dark red curve shows a similar trend; however, the final yield is more than doubled and reaches a significantly higher value of 9.98.

To further explore the relationship between optical absorption and catalytic performance, the product yield at 10 seconds is plotted against the total measured optical absorption in Figure 5b. A pronounced enhancement in catalytic yield is observed upon switching the laser polarization from TE to TM, which correlates with the gradually increased plasmonic resonance strength as evidenced by the absorption spectra. Additionally, measurements under linearly polarized light at 60° (green data point in Figure 5b) also show a considerable increase in absorption to 65%, accompanied by a corresponding yield enhancement to 6.1. Similarly, for the 30° polarization case, the yield reaches 8.6—approaching the value obtained under TM (0°) polarization. The catalytic reactivity is continuously modulated by gradually shifting the light polarization, enabling controllable reaction yields.

These findings collectively demonstrate that the yield of the MB photocatalytic reaction can be dynamically modulated by actively tuning the plasmonic resonance strength through polarization control of the incident light on the Au-TiO$_2$ NP metasurface. The total absorption analysis reveals the rate of resonance enhancement at the target wavelength and provides direct insight into the catalytic reaction activity. The metasurface is engineered to support strong near-field and hot-electron generation rate under TM polarization, aligning with the electronic transition of MB (*i.e.*, the HOMO-LUMO gap), thereby significantly enhancing the photocatalytic yield. Conversely, polarization switching results in misalignment between the plasmonic and molecular resonances due to a redshift of the plasmonic peak with increasing NP radius, reducing the catalytic efficiency. Figure 5b demonstrates a proportional relationship between the catalytic yield and the measured optical absorption, highlighting the capability to modulate reactivity to generate particular product yield through tunable optical properties. In the current model reaction, a slight mismatch remains between the peak resonance and the molecular resonance, due to fabrication uncertainties. With improved alignment, a more significant difference in catalytic yield is anticipated.

More broadly, elliptical Au-TiO$_2$ NP metasurfaces provide an effective reaction environment which not only adsorbs the reactants[34,35] but also plasmonically tailors the photocatalytic performance. The photocatalytic yield is actively manipulated by modulating the strength of the LSPR. Reactant molecules are typically adsorbed onto the localized surface of gold nanostructures, where they interact with injected photons or hot electrons. To achieve active control over such reactions, metasurfaces must exhibit tunability through a particular structural design. Here, we employ elliptical Au-TiO$_2$ NP arrays and achieve tunability in the plasmonic resonance strength through a highly compact, asymmetric morphological design, while maintaining identical illumination parameters (*e.g.*, wavelength, power, objective lens and time duration). The only variable is the polarization state of the incident light.

Under TM polarization, the wavelength-dependent plasmonic resonance selectively catalyzes reactions whose molecular resonances align with the resonance peak. By switching the polarization of the illumination source from TM to TE, the resonance wavelength can be shifted toward longer wavelengths, while simultaneously suppressing the resonance intensity at the original peak due to the orthogonal polarization. The optical and catalytic behaviors exhibit a proportional relationship in catalytic yield. Consequently, under TE polarization, the metasurface is expected to catalyze reactions aligned with longer-wavelength resonances. Meanwhile, other reactions associated with shorter-wavelength resonances, previously active under TM mode, are suppressed due to reduce plasmonic resonance. This phenomenon reveals the potential of this strategy in selectively catalyzing particular reactions aligned with different molecular resonance energies.

Wavelength-dependent plasmonic resonance enables the selective enhancement of specific photocatalytic reactions while suppressing those that are misaligned with the effective resonance wavelength range. The polarization-dependent tunability of the resonance peak further offers a flexible strategy for continuous controlling reaction yields. Desired yields can be achieved through the continuous tuning of optical property, which are linearly correlated with catalytic performance. This feature allows the targeting of two distinct photocatalytic reactions, each corresponding to a different wavelength range and selectively activated by orthogonal polarizations. A single metasurface configuration thus provides a compact and versatile platform with significant potential for managing complex, multi-branched photocatalytic processes.

CONCLUSION

We demonstrate the active control of photoreactivity using a polarization-dependent metasurface composed of Au-TiO$_2$ NP arrays. Through accurate morphology design, we engineered asymmetric elliptical Au nanodisks on top of TiO$_2$ NPs to support polarization-sensitive plasmonic resonances at different wavelengths under the two orthogonal polarization states. At the target wavelength of 633 nm, the metasurface exhibits a strong plasmonic resonance under TM polarization, achieving 89% absorption and enhanced near-field confinement that promotes hot-electron generation and efficient interaction with reactants. This drives the N-demethylation of MB, quantified *via* real-time SERS by integrating the ~480 cm$^{-1}$ Raman peak, yielding 9.98 after 10s under 3.4 mW TM polarized illumination. In contrast, TE polarization causes a redshift in the Au nanodisk resonance, reducing absorption to 58% and confining the near field effects of Au nanodisks. As a result, plasmonic resonance becomes misaligned with the MB HOMO-LUMO gap, suppressing reactivity and reducing the yield to 4.7. These results highlight that by tuning the plasmonic resonance strength through structural asymmetry, we can precisely control resonance-driven photocatalytic reactions with only switching the polarization state of the incident light. This approach provides a compact and adaptable solution for continuous modulating chemical reactivity, holding substantial promise for the selective catalysis of multi-pathway chemical reactions achieving desired reactivity.

METHODS

**Finite element method simulation.** The far-field optical properties of the Au-TiO$_2$ NP array system were modeled with the finite element method (FEM) using COMSOL Multiphysics 5.6. The refractive index of TiO$_2$ in this model was based on measurements of a

100 nm thin film deposited under similar condition to the experimental samples, using an ellipsometer (JA Woollam M-2000D) covering wavelengths from 200 to 1700 nm. Floquet periodic boundary conditions were applied to all four surrounding surfaces to simulate a periodic array of nanopillars. The incident light was vertically coupling to the nanostructure, introduced through a port with wave excitation. The polarization is varied from TM to TE with step polarization angle of 10°. The total absorption was determined using the equation 1−T−R, where T represents total transmittance, and R corresponds to total reflectance. The optical response of the metasurface was examined as a function of the NPs radius in x and y-direction, height of Au NDs and period of Au-TiO$_2$ NPs. The effect of TiO$_2$ NP tapering on the optical properties is also investigated. Additional simulation results can be found in the SI.

**Sample preparation.** The Au-TiO$_2$ NP samples were prepared on the 1 mm glass slide substrate. The gold mirror layer was deposited using electron-beam (E-beam) evaporation (Temescal BJD-2000) under pressure of $1 \times 10^{-5}$ torr. The deposition rate was $1 \ Å/s$ and final thickness was 200 nm. The anatase TiO$_2$ layer was sputtered onto the Au mirror at an approximate deposition rate of 0.17 Å /s with the chamber pressure maintained below $4 \times 10^{-6}$ Torr. The layer thickness is 100 nm. The pattern was defined by electron-beam lithography (Raith 150) under the base dose $50 \ \mu C/cm^2$ and dose factor from 2.5 to 3.2. Bilayer PMMA was applied as the photoresist, which contains PMMA 495 A2 and PMMA 950 A4. After that, a 14 nm Au and 30 nm Ni were deposited *via* E-Beam evaporation under a pressure of $1 \times 10^{-5}$ Torr followed by photoresist lifted off in acetone for over 4 hours. The nickel nanodisks were employed as the mask for plasma etching. The TiO$_2$ film was etched with Ar and CHF$_3$ plasma for 6.5 minutes. The Ni masks were finally removed with FeCl$_3$ for 1 minute.

**Micro-spectrophotometer.** The absorption spectra were calculated with 1-R-T, where R and T are the reflectance and transmission of models, respectively. A Xenon Arc Light Source (Thorlabs SLS401) provided illumination across the full visible spectrum, covering wavelengths from 300 to 800 nm. The polarization of the light is adjusted with a rotatable linear polarizer (Thorlabs LPVISE100-A). The measurement area was defined using a square aperture, matched to the array size of 100×100 $\mu m$. A 10× objective lens (Olympus Mplan N) was used to focus the light onto the sample, and the reflected and transmitted light was collected within an acceptance angle of 22°, which was then directed to a fiber spectrometer (StellarNet Inc. BLACK-Comet-SR). The optical properties were measured under the varying polarization angle from 0° to 90°, with a step of 10°. To minimize random noise, the spectrum was integrated over 100 seconds and averaged over five measurements.

**Surface enhanced Raman spectroscopy.** The Renishaw inVia Reflex Raman spectroscopy was employed to measure the Raman spectra of samples under different polarizers. The central wavelength of the laser is 633nm and the light is linearly polarized. The objective lens is Olympus LMPLFLN20X. The grating grooves are 1200 lines/mm. During the SERS measurement, the Raman signals were collected in a 10s period with a 0.1s time interval. The polarization angle was adjusted by rotating the sample under 0°, 30°, 60°, 90° in polarization angle.

ASSOCIATED CONTENT

**Supporting Information.** The supporting information is shown at *Polarization-Sensitive Au-TiO$_2$ Nanopillars for Tailored Photocatalytic Activity_SI.pdf.*

We present the supplementary simulation results for the model reaction, N-demethylation of Methylene blue degradation in Figure S1, revealing the processing of the resonance driven reaction. (PDF). The absorption contributed by plasmonic nanostructures are studied by FEM simulations shown in Figure S2. Additionally, sample preparation processes of Au-TiO$_2$

NP arrays are revealed in Figure S3 and analyzed in morphological defects in Figure S4. Figure S5 illustrates the SEM images of Au-TiO$_2$ NP arrays with various lithography electron beam dose. Following, the optical properties of these samples are shown in Figure S6, reflecting the morphological engineering of nanopillars in radius. And the substrate absorption is shown in Figure S7 as referencing. Finally, in SERS measurement, Raman spectrums are shown in Figure S8-S11 analyzed at specific peak wavenumbers in Table S1. (PDF)

AUTHOR INFORMATION

**Corresponding Authors**

**Christin David** - Institute of Solid State Theory and Optics, Friedrich-Schiller-Universität Jena, 07743 Jena, Germany; University of Applied Sciences Landshut, Am Lurzenhof 1, 84036 Landshut, Germany. Email: christin.david@uni-jena.de

**Fiona J. Beck** - School of Engineering, Australian National University, ACT 2601, Australia. Email: fiona.beck@anu.edu.au

**Ning Lyu** - Institute of Solid State Theory and Optics, Friedrich-Schiller-Universität Jena, 07743 Jena, Germany; School of Engineering, Australian National University, ACT 2601, Australia. Email: ning.lyu@uni-jena.de

**Authors**

**Anjalie Edirisooriya** - School of Engineering, Australian National University, ACT 2601, Australia.

**Zelio Fusco** - School of Engineering, Australian National University, ACT 2601, Australia.

**Dawei Liu** - Australian Research Council Centre of Excellence for Transformative Meta-Optical Systems, Department of Electronic Materials Engineering, Research School of Physics, The Australian National University, Canberra ACT 2601, Australia; Institute of Applied Physics, Abbe Centre of Photonics, Friedrich Schiller University Jena, Albert-Einstein-Str. 15, Jena 07745, Germany.

**Lan Fu** - Australian Research Council Centre of Excellence for Transformative Meta-Optical Systems, Department of Electronic Materials Engineering, Research School of Physics, The Australian National University, Canberra ACT 2601, Australia.

**Author Contributions**

The manuscript was written through contributions of all authors. All authors have given approval to the final version of the manuscript.

**Funding Sources**

The project is funded by International Research Training Group IRTG 2675 (GEPRIS 437527638).

ACKNOWLEDGMENT

This work is supported by the German Research Foundation DFG (Deutsche Forschungsgemeinschaft) through funding of the International Research Training Group IRTG 2675 (GEPRIS 437527638). The authors acknowledge access to NCRIS facilities (ANFF-ACT Node) at the Australian National University.

ABBREVIATIONS

MB, methylene blue; LSPR, localized surface plasmon resonance; TE, transverse electric; TM, transverse magnetic; BIC, bound states in the continuum; Au-TiO$_2$ NPs, Au-TiO$_2$ nanopillars; Au ND, Au nanodisk; SERS, surface-enhanced Raman spectroscopy; HOMO, highest occupied molecular orbital; LUMO, lowest unoccupied molecular orbital; PMMA, polymethyl methacrylate; EBL, electron-beam lithography; SEM, scanning electron microscope; FEM, finite element method.

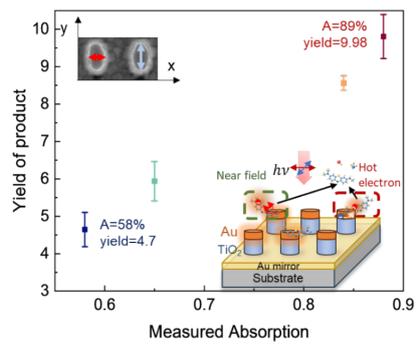

For Table of Contents Only

Supporting information

# Polarization-Sensitive Au–TiO$_2$ Nanopillars for Tailored Photocatalytic Activity

*Ning Lyu [a b\*], Anjalie Edirisooriya [b], Zelio Fusco [b], Dawei Liu [c d], Lan Fu [c], Fiona J. Beck [b\*], Christin David [a e\*]*

a. Institute of Solid State Theory and Optics, Friedrich-Schiller-Universität Jena, 07743 Jena, Germany

b. School of Engineering, Australian National University, ACT 2601, Australia

c. Australian Research Council Centre of Excellence for Transformative Meta-Optical Systems, Department of Electronic Materials Engineering, Research School of Physics, The Australian National University, Canberra ACT 2601, Australia

d. Institute of Applied Physics, Abbe Centre of Photonics, Friedrich Schiller University Jena, Albert-Einstein-Str. 15, Jena 07745, Germany

e. University of Applied Sciences Landshut, Am Lurzenhof 1, 84036 Landshut, Germany

\* Corresponding authors: christin.david@uni-jena.de (C. David); fiona.beck@anu.edu.au (F.J. Beck); ning.lyu@uni-jena.de (N. Lyu)

# N-DEMETHYLATION OF METHYLENE BLUE (MB)

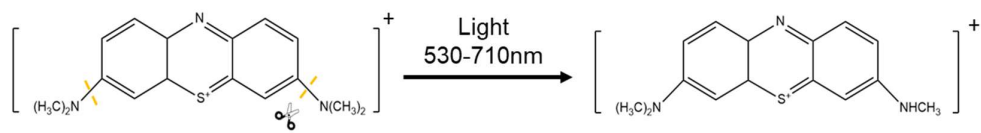

**Figure S1**: Depiction of the photocatalytic reaction of N-demethylation of methylene blue.

In this reaction, C-N bonds were cleavage forming the product, N-demethylated derivatives. This process can be catalyzed with light in wavelength range ~530 – 710 nm, which is characterized with the photoluminescence spectrum. The end product of the N-demethylation is thionine.[1] The theoretical studies reveal that MB molecules are able to adsorb on the surface of gold via weak dispersion force.[2,3]

# SIMULATION RESULTS OF ELLIPTICAL AU-TIO$_2$ NANOPILLAR (AU-TIO$_2$ NP) ARRAY MODEL

**Finite element method simulation.**

The near-field and far-field optical properties of the Au-TiO$_2$ NP array system were modeled with the finite element method (FEM) using *COMSOL Multiphysics 5.6*. Simulation method can be found in the Method section.

The simulation results of the Au-TiO$_2$ nanopillar array reveal the absorption spectra under both transverse magnetic (TM) and transverse electric (TE) light illumination. The model consists of an array of elliptical TiO$_2$ nanopillars, each capped with a gold nanodisk (ND). In Figure S2, the structure exhibits asymmetry, with radius of 35 nm along the x-axis and 65 nm along the y-axis, leading to distinct absorption peaks at 630 nm and 850 nm, respectively.

Under TM polarization, the absorption peak at 630 nm is primarily due to the Au ND, which contributes 96% of the total absorption at that wavelength. For TE polarization, the peak absorption occurs at 850 nm, where the absorption of Au ND accounts for almost the entire absorption of 57% at that point. These results indicate a significant enhancement of plasmonic resonance at the respective peak wavelengths for both polarizations.

However, under TE illumination at shorter wavelengths, the absorption by the Au ND decreases. Specifically, close to the target wavelength at 630 nm, 18 percentage points of the total 27% absorption are attributed to the Au ND. This reduction in the contribution leads to a weaker plasmonic resonance, highlighting the polarization-dependent behaviour of the structure due to its asymmetric geometry.

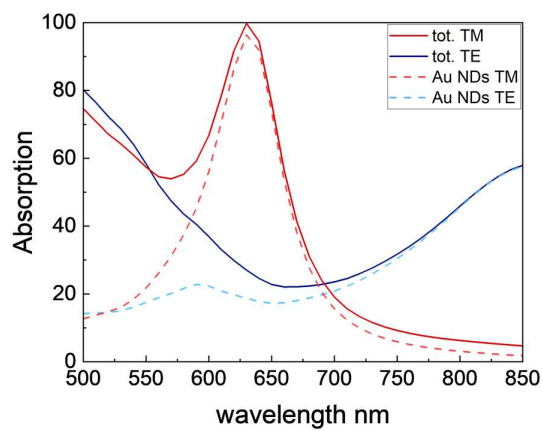

**Figure S2**: Simulated absorption of a Au-TiO$_2$ NP array with $R_x$=35 nm and $R_y$=60 nm. Under TM polarization, the total absorption (red solid curve) is primarily provided by the absorption of Au nanodisks (NDs) on top of the TiO$_2$ NP (red dashed curve) at the peak wavelength 630 nm. In contrast, in the TE mode, the total absorption (blue solid curve) is only partially due to the absorption of Au ND (blue dash curve) at this wavelength. Instead, the peak wavelength at 850 nm is provided by the plasmonic resonance.

# FABRICATION PROCESS OF ELLIPTICAL AU-TIO₂ NP ARRAYS

**Sample preparation.**

The Au-TiO₂ NP samples were prepared on the 1 mm glass slide substrate. The gold mirror layer was deposited using electron-beam (E-beam) evaporation (Temescal BJD-2000) under pressure of $1 \times 10^{-5}$ torr. The deposition rate was 1 Å/$s$ and final thickness was 200 nm. The anatase TiO₂ layer was sputtered onto the Au mirror at an approximate deposition rate of 0.17 Å /s with the chamber pressure maintained below $4 \times 10^{-6}$ Torr. The layer thickness is 100 nm. The pattern was defined by electron-beam lithography (Raith 150) under the base dose $50\ \mu C/cm^2$ and dose factor from 2.5 to 3.2. Bilayer PMMA was applied as the photoresist, which contains PMMA 495 A2 and PMMA 950 A4. After that, a 14 nm Au and 30 nm Ni were deposited *via* E-Beam evaporation under a pressure of $1 \times 10^{-5}$ Torr followed by photoresist lifted off in acetone for over 4 hours. The nickel nanodisks were employed as the mask for plasma etching. The TiO₂ film was etched with Ar and CHF₃ plasma for 6.5 minutes. The Ni masks were finally removed with FeCl₃ for 1 minute.

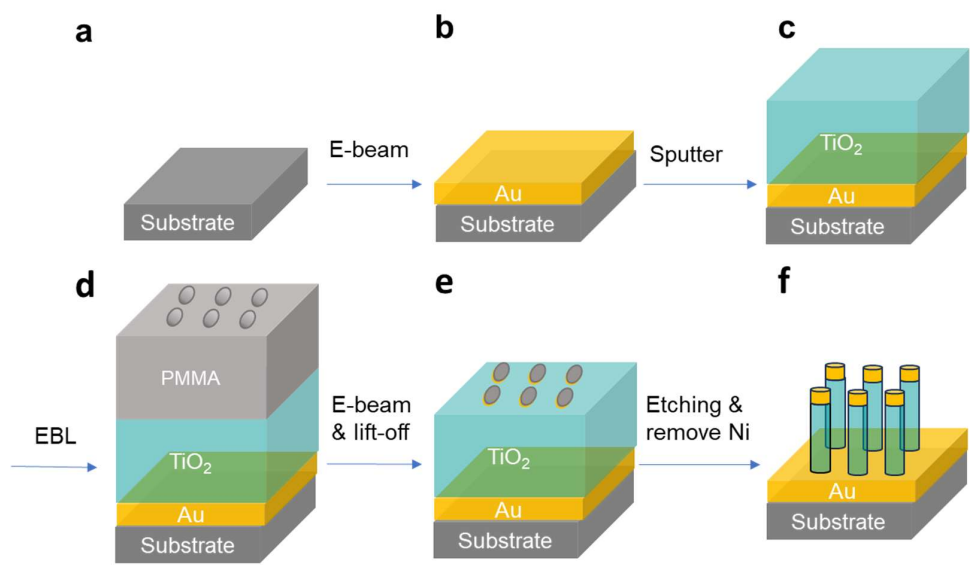

**Figure S3:** Fabrication process of the Au-TiO₂ NP array. a. The substrate is SiO₂ glass slides. b. A Au mirror film is deposited with an electrical-beam evaporator for 200 nm. c. 100 nm

anatase TiO$_2$ is sputtered onto the gold film with the thickness controlled by the duration period. d. Elliptical patterns are written with electron-beam lithography (EBL) with the bilayer PMMA photoresist. After exposure, the sample develops with MIBK: IPA 1:3. e. 14 nm Au and 30 nm Ni are deposited with the electrical-beam evaporator. Ni acts as the mask, followed by an etching step. The photoresist is removed when exposed to acetone for 4 hours. f. TiO$_2$ film etching with CHF$_3$ and Ar plasma forms the nanopillar array. The Ni mask is finally removed with 5% FeCl$_3$ for 1 min.

**Simulation results of Au-TiO$_2$ NP array with defects**

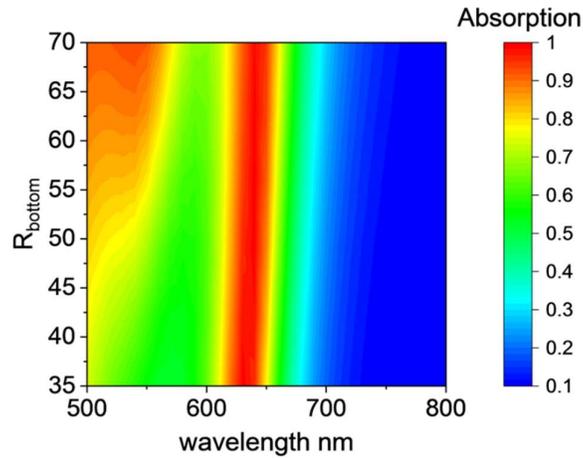

**Figure S4**: Total absorption of a tapered TiO$_2$ NP array as a function of the bottom NP radius. The plasmonic peak wavelength shifts slightly from 632 to 637 nm and the bandwidth broadens from 49 to 56 nm.

Defects in the Au-TiO$_2$ NP array are observable in the scanning electron microscope (SEM) images. The TiO$_2$ nanopillars exhibit a tapered radius along both the x- and y-axes. In the simulation, the nanopillars are modelled with top radius matching the design parameters 35 nm in the x-direction and 60 nm in the y-direction. The radius gradually increases along the height of the nanopillar, maintaining the same magnitude in both perpendicular directions.

To account for this tapering, the bottom radius in the x-direction is swept in the simulation from 35 nm to 70 nm. The magnification expansion is applied equally in both directions, consistent with the observations in the SEM images. Accordingly, the bottom radius in the y-direction varies from 60 nm to 95 nm in the simulation.

The total absorption under TM-polarized illumination is calculated to investigate the optical properties at the target wavelength. The absorption spectra are plotted as a function of the bottom radius along the x-direction. A slight redshift in the peak wavelength from 632 nm to 637 nm is observed as the bottom radius doubles, though this shift is minimal. However, a noticeable broadening of the bandwidth occurs, increasing from 49 nm to 56 nm.

**Scanning electron microscope (SEM) image of the sample with various EBL doses**

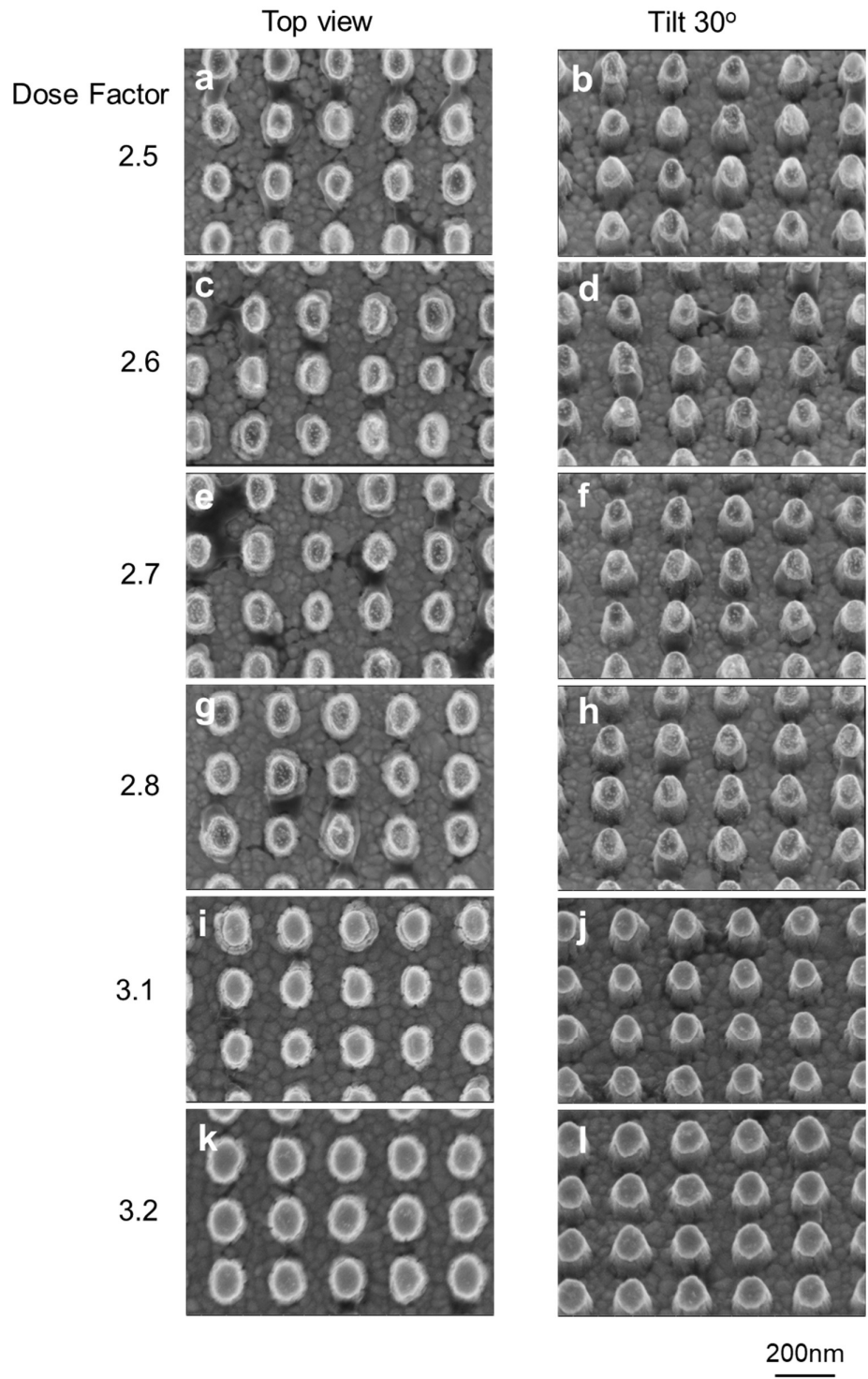

**Figure S5**: SEM image of the Au-TiO$_2$ NP arrays prepared with increasing exposure doses of EBL. The nanopillar radius is enlarged in both x- and y-direction with higher dose. The reference dose is 50 $\mu C/cm^2$. The radius increases with the dose[4] and has the average parameters $R_x$ = 27 nm, $R_y$ = 49 nm (dose 2.5); $R_x$ = 31 nm, $R_y$ = 50 nm (dose 2.6); $R_x$ = 34 nm, $R_y$ = 52 nm (dose 2.7); $R_x$ = 36 nm, $R_y$ = 55 nm (dose 2.8); $R_x$ = 40 nm, $R_y$ = 56 nm (dose 3.1); $R_x$ = 42 nm, $R_y$ = 58 nm (dose 3.2).

SEM images of the Au-TiO$_2$ NP array are shown for increasing EBL doses, ranging from 125 to 160 $\mu C/cm^2$, captured in both top views and 30° tilt views. These images are used for morphological analysis. The TiO$_2$ NPs exhibit a tapered shape, with the radius increasing uniformly along both the x- and y-axes. This tapering results in a slight broadening of the absorption spectrum bandwidths, as shown in Figure S4.

Additionally, the average radius of the Au-TiO$_2$ NPs increases with higher EBL doses. The corresponding shifts in optical properties due to this radius variation are discussed in the following section.

# OPTICAL PROPERTIES OF AU-TIO$_2$ NP ARRAYS

## Optical properties of Au-TiO$_2$ NP arrays with various nanopillar parameters

The optical properties of the Au-TiO$_2$ NP array were measured with micro-spectrophotometer under various polarization modes TM and TE. As shown in the SEM images in Figure S5, the geometry of the Au-TiO$_2$ NP structure, which is affected by the EBL dose, determines the peak wavelength under both modes. By appropriately selecting the EBL dose, the absorption peak can be finely tuned to match the target resonance wavelength of MB reactions.

For both polarization modes, an increase in nanopillar radius leads to redshift in the peak wavelength. For example, under TM mode, the radius in the x-direction ($R_x$) increases from 610 nm to 655 nm as the average NP radius grows from 27 to 42 nm. Similarly, under TE mode, the peak wavelength shifts from 730 nm to 810 nm as the radius in the y-direction ($R_y$) increases from 49 to 58 nm. These redshifts are attributed to changes in the geometry of the Au NDs, highlighting their role in enhancing plasmonic resonance.

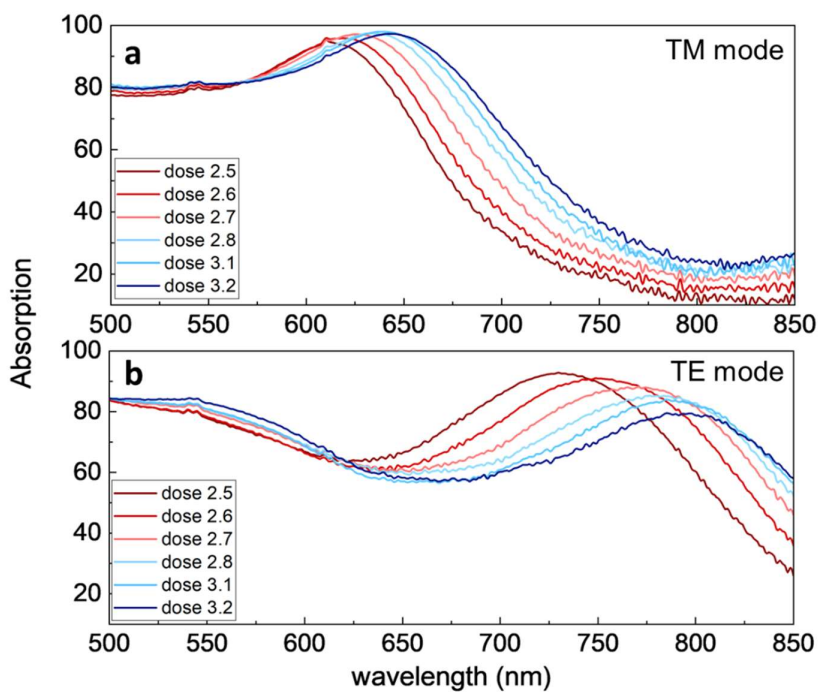

**Figure S6**: Measured absorption of Au-TiO$_2$ NP arrays with various dose levels of EBL. The reference dose is 50 $\mu C/cm^2$. The radius increases with the dose[4] and has the average parameters R$_x$ = 27 nm, R$_y$ = 49 nm (dose 2.5, dark red curves); R$_x$ = 31 nm, R$_y$ = 50 nm (dose 2.6, red curves); R$_x$ = 34 nm, Ry = 52 nm (dose 2.7, pink curves); Rx = 36 nm, Ry = 55 nm (dose 2.8, light blue curves); Rx = 40 nm, Ry = 56 nm (dose 3.1, blue curves); R$_x$ = 42 nm, R$_y$ = 58 nm (dose 3.2, dark blue curves). The peak notably redshifts under both TM and TE mode.

Additionally, reference samples were used to evaluate the contribution of the Au NDs. The sample with an Au mirror on the substrate exhibits strong absorption below 550 nm and high reflectance in the 600 - 900 nm range designed as the mirror layer. The black curve in Figure S7 represents the total absorption spectrum of a sample with a 100 nm TiO$_2$ film deposited on an Au mirror. This TiO$_2$ layer forms a semi-open cavity, resulting in a peak absorption at 560 nm, though with a relatively low intensity of only 36%. In contrast, the Au ND-TiO$_2$ nanopillar array, as shown in Figure S6, significantly enhances absorption through plasmonic resonance.

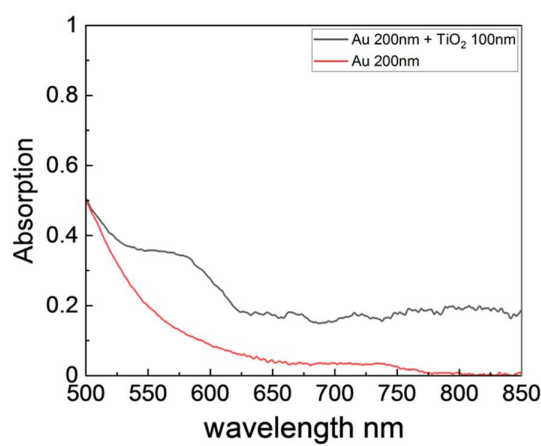

**Figure S7**: Measured absorption spectra of the 200 nm Au mirror substrate and the sample after deposition of a 100 nm anatase TiO$_2$ film. The Au substrate exhibits constant absorption in the 550 - 850 nm wavelength range. Following the TiO$_2$ deposition, the absorption increases slightly to 10% due to the formation of a semi-open cavity[5].

# CATALYTIC CHARACTERIZATION WITH SURFACE ENHANCED RAMAN SPECTROSCOPY (SERS)

**Specific peak of Surface enhanced Raman spectrums.**

The specific peak of MB and products are shown in the table below refer to the references[1,2,6].

**Table 1** Raman spectrum peaks and band assignment for MB and N-Demethylated Derivatives

| Raman shift cm$^{-1}$ | Band assignment |
|---|---|
| 446 | C-N-C skeletal deformation |
| 479 | skeletal deformation mode of thionine |
| 501 | C-N-C skeletal deformation |
| 610 | C-S-C skeletal deformation |
| 772 | C-H in-plane bending |
| 804 | NH$_2$ rocking vibration of thionine |
| 1390 | C-N symmetric stretching |
| 1426 | C-N asymmetric stretching |
| 1434 | C-N asymmetric stretching |
| 1622 | C-C ring stretching |

**Surface enhanced Raman spectroscopy (SERS) results of Au-TiO$_2$ NP arrays measured under different polarization states**

The progression of the chemical reaction was monitored using in-situ surface-enhanced Raman spectroscopy (SERS) measurements at 0.1 s intervals over a 10 s period. A linearly polarized 633 nm laser with a power of 3.4 mW, corresponding to the catalytic wavelength for MB, was used for excitation. The spectra were normalized to the benzene ring peak at ~1600 cm$^{-1}$. The reaction product was identified and quantified by tracking its characteristic Raman peak at approximately 480 cm$^{-1}$.

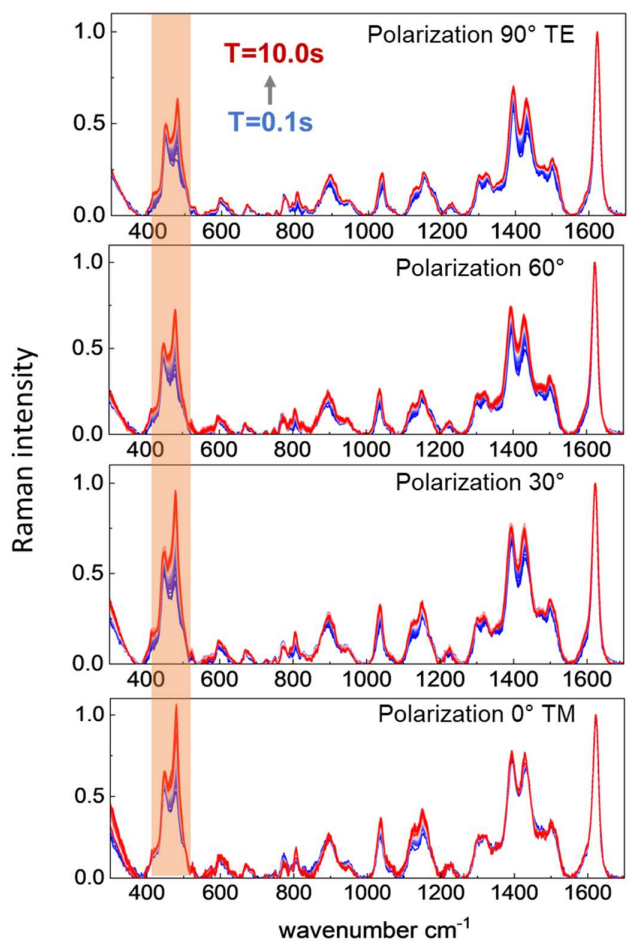

**Figure S8**: Normalized Raman spectrum under 0° (TM), 30°, 60°, 90° (TE) polarization. The Raman peak of the designed Au-TiO$_2$ NP arrays (with average $R_x$= 35 nm and $R_y$= 60 nm) monitors the reaction in a 10 s period with 0.1 s intervals from dark blue curves to dark red curves. The orange area shows the area of interest[6], the Raman peak wavenumber range 420 to 510 cm$^{-1}$ for curve fitting.

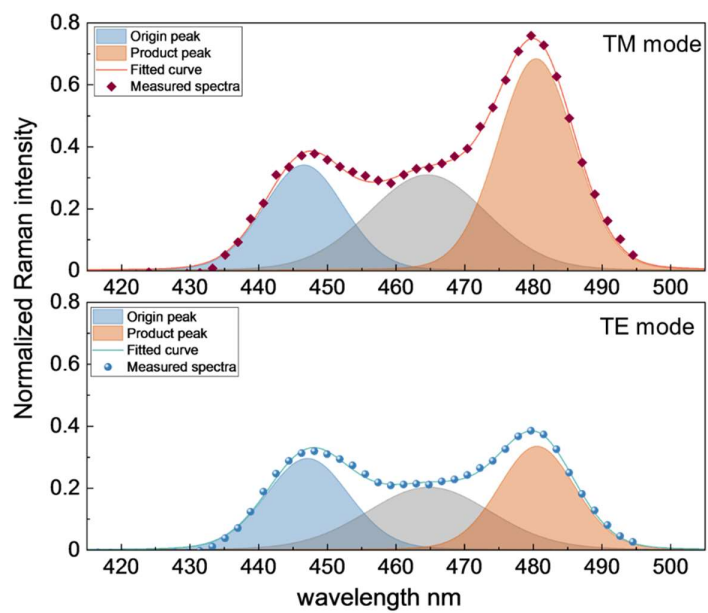

**Figure S9**: Peak fitting of the Au-TiO$_2$ NP arrays at 10 s under TE (bottom) and TM (top) mode laser excitation. The original (blue curves) and product (orange curves) peaks are located at approximately 445 cm$^{-1}$ and 480 cm$^{-1}$, respectively[2,6]. The fitted curves (cyan and red) include three peaks and show good agreement with the measured spectra (blue and dark red dots), with residual errors below 4%.

**SERS results of Au-TiO$_2$ NP arrays measured under different laser power**

The Raman spectrums were used to monitor the MB reaction under varying laser powers, ranging from 0.01 mW to 0.52 mW. As the laser power increases, the plasmonic resonance of the Au NPs is enhanced, leading to a more pronounced change at the characteristic product peak, indicating an increase in product yield. Notably, a significant product peak increase is observed when the laser power exceeds 0.1 mW.

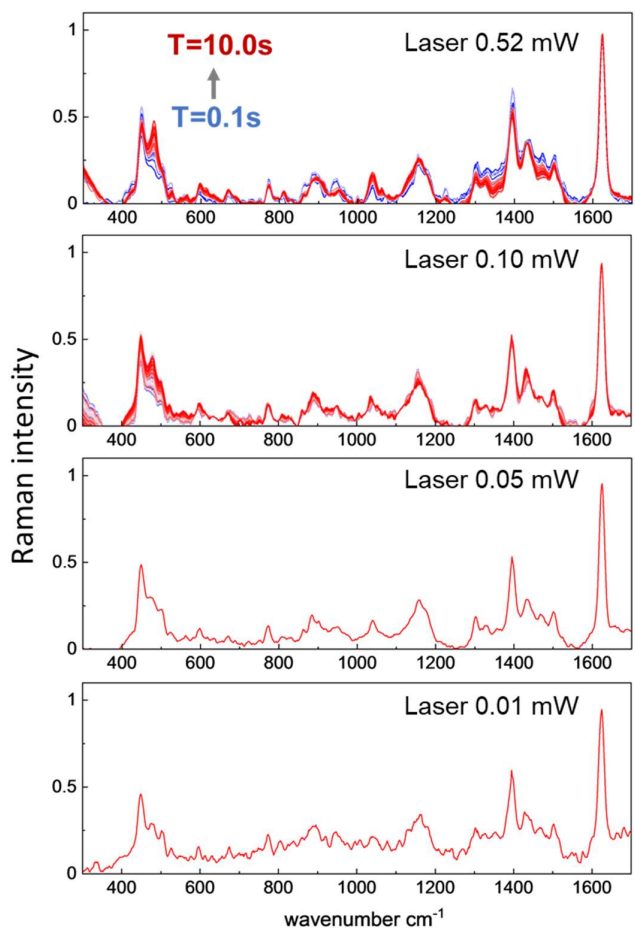

**Figure S10**: Normalized Raman spectrum under linear 633 nm laser with varying power 0.01 mW, 0.05 mW, 0.10 mW and 0.52 mW. The Raman peak of the designed Au-TiO$_2$ NP arrays ($R_x$= 35 nm and $R_y$= 60 nm) is characterized with a 10 s period interval in 0.1 s. The product peak 480 cm$^{-1}$ change is negligible when the laser power remains under 0.05 mW and increases with power.

Sample stability was evaluated at five different positions under various polarization angles (0°, 30°, 60°, and 90°), as shown in Figure S11. Measurements were conducted under consistent conditions using a 633 nm laser at 3.4 mW to ensure consistency of measurement conditions. The standard deviation of the product yield at 10 seconds, as presented in Figure 5B, was calculated based on the results from these five measurements.

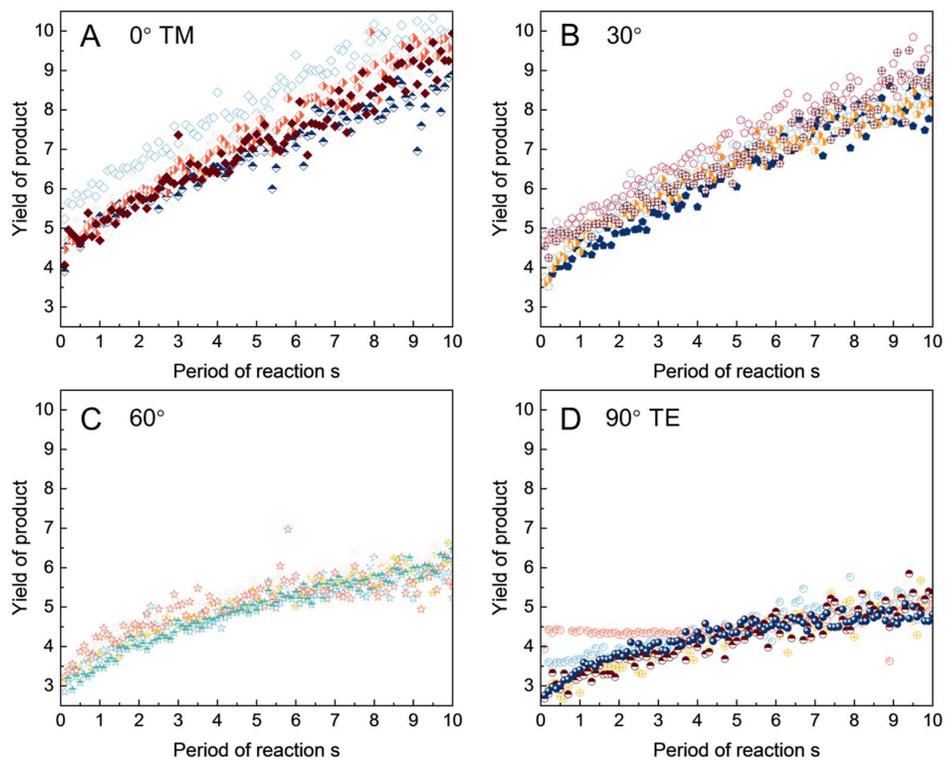

**Figure S11**: Reaction yield of Au-TiO$_2$ NP arrays as a function of processing time under four different polarization states: 0° (A), 30° (B), 60° (C), and 90° (D). For each polarization, five measurements were taken at different positions within the array, demonstrating the stability and reproducibility of the results.